%% file: sparticleB-L.tex
\documentclass[12pt,letterpaper]{article}

\usepackage{epsfig}
\usepackage{amsmath}
\usepackage{verbatim}
\usepackage{psfrag}
\usepackage{bm}
\usepackage{bbm}
\usepackage[square,comma,sort&compress]{natbib}

\begin{document}

\vskip .5cm

\begin{center}

{\Large\bf The Effect of the Sparticle Mass Spectrum on the
\\
\vskip.3cm
Conversion of $B-L$ to $B$}

\vskip 1cm

Daniel~J.~H.~Chung,$^{\textnormal{\it a,b}}$
Bj\"orn~Garbrecht,$^{\textnormal{\it a}}$
and Sean~Tulin$^{\textnormal{\it c}}$

\vskip .4cm

{\it
$^{\textnormal{\it a}}$University of Wisconsin-Madison,
Department of Physics,\\
1150 University Avenue, Madison, WI 53706, USA
}
\vskip .2cm
{\it 
$^{\textnormal{\it b}}$School of Physics,
Korea Institute for Advanced Study, 207-43,\\
Cheongnyangni2-dong, Dongdaemun-gu, Seoul 130-722, Korea
}

\vskip .2cm

{\it
$^{\textnormal{\it c}}$California Institute of Technology,
1200 E.~California~Blvd.,\\
Pasadena, CA 91125, USA
}

\end{center}

\vskip .5cm

\begin{abstract}
In the context of many leptogenesis and baryogenesis scenarios, $B-L$
(baryon minus the lepton number) is converted into $B$ (baryon number)
by non-perturbative $B+L$ violating operators in the ${\rm SU}(2)_L$ sector.
We correct a common misconversion of $B-L$ to $B$ in the literature in
the context of supersymmetry.  More specifically, kinematic effects
associated with the sparticle masses can be generically important
(typically a factor of $2/3$ correction in mSUGRA scenarios), and in some cases, it may
even flip the sign between $B-L$ and $B$.  We give explicit formulae
for converting $B-L$ to $B$ for temperatures approaching the
electroweak phase transition temperature from above.  Enhancements of
$B$ are also possible, leading to a mild relaxation of the reheating
temperature bounds coming from gravitino constraints.
\end{abstract}

\section{Introduction}

In many leptogenesis and baryogenesis scenarios
\cite{Fukugita:1986hr,Buchmuller:2004nz,Pilaftsis:2003gt,Davidson:2008bu,Riotto:1999yt,Dine:2003ax,Dolgov:1991fr,Luty:1992un},
conserved charges associated with baryon minus the
lepton number ($B-L$ or $B/3-L_i$) are generated by out-of-equilibrium
dynamics. Subsequently, $B-L$ or $B/3-L_i$ is converted to the
observed baryon number $B$ by non-perturbative ${\rm SU}(2)_L$ operators, which are in equilibrium until the time of the electroweak phase transition.  Many
papers
\cite{Kuzmin:1987wn,Khlebnikov:1988sr,Harvey:1990qw,Dreiner:1992vm,Davidson:1994gn,Khlebnikov:1996vj,Laine:1999wv,Burnier:2005hp}
have thus investigated the relationship between $B-L$ or $B/3-L_i$ and
$B$.  To leading order in perturbative expansion, the relationship
between $B-L$ and $B$ that is used in the literature is that of Ref.~\cite{Harvey:1990qw}.  For example, in MSSM-like situations with two
Higgs doublets, assuming a strong first order phase transition, the formula 
\begin{equation}
B=\frac{8}{23} (B-L)
\label{eq:typical}
\end{equation}
is often used to convert the $B-L$ generated by thermal leptogenesis into the
observed baryon asymmetry~\cite{Davidson:2008bu}.  In this article, we point out that
kinematics associated with beyond the Standard Model (SM) particles can
significantly change this kind of formula such that the usage of
Eq.~(\ref{eq:typical}) is incorrect.  Although ``typical'' changes in
the formula are modest, in some cases the change may be an order of
magnitude or may even lead to a sign flip between $B$ and $B-L$, even
within just the Minimal Supersymmetric Standard Model (MSSM) particle content.  In the case that the constant
of proportionality between $B$ and $B-L$ is increased from the
standard value, the well known reheating temperature bounds associated
with gravitinos \cite{Kawasaki:2004yh,Khlopov:1984pf,Ellis:1984eq} may
be mildly relaxed.

Although the effects of particle content and kinematics on the
relationship between $B$ and $B-L$ have been discussed to some extent
before \cite{Davidson:1994gn,Dreiner:1992vm}, we contribute the following,
using the MSSM spectrum to make the illustration concrete:
\begin{enumerate}
\item We give a closed form expression relating $B$ to $(B-L)$ or $(B/3-L_i)$
  that includes all the sparticle kinematic effects.
\item We point out that there are situations where $B$ and $B-L$ can
  have opposite signs, or $B$ can be larger than some of the typical
  values used in the literature.
\item Even when some sparticles are heavier than the temperature,
  equilibrium can be maintained, although its effect on the system
  will become smaller as the number density becomes Boltzmann
  suppressed.
\item Even if gauginos are extremely heavy, ``superequilibrium'' --- namely, chemical equilibrium between a particle and its superpartner --- is maintained through Yukawa interactions.
\end{enumerate}
Although the latter two points have been discussed in the context of
electroweak baryogenesis in \cite{ourbgenesis}, they are also
important for lepto-/baryogenesis scenarios involving the conversion from
$B-L$ to $B$.

The order of presentation will be as follows.  In the next section we
briefly discuss the intuition for the $B-L$ to $B$ conversion.  In
Sec.~\ref{sec:equilib}, the formulae relating $B-L$ (or $B/3-L_i$) to
$B$ are derived.  Sec.~\ref{sec:discuss} discusses the range of
possible numerical effects coming from the derived formulae.  We
then conclude with a brief summary.

\section{\label{sec:intuition}Intuition}

We briefly sketch the main physical point of this work such that the reader may gain some intuition.  Suppose, at some early time, with temperature $T \gg 100$ GeV, $(B-L)$ is generated. Although $(B-L)$ is approximately conserved, individually $B$ and $L$ are affected by the non-perturbative ${\rm SU}(2)_L$ operator; this process is in equilibrium and leads to the condition that
\begin{equation}
3 B_L \approx - L_L \;,
\label{eq:3bplusL}
\end{equation}
where the subscript $_L$ refers to baryon and lepton number in left
handed fermions.  (In Eq.~\eqref{eq:3bplusL}, we have assumed
that the thermal masses of quarks and leptons are light compared to
the temperature.  Recall we are considering a situation before the
electroweak phase transition, when SM fermions have no
Higgs-vev-induced masses.)  It is important to note that
Eq.~(\ref{eq:3bplusL}) is valid even though for each sphaleron
transition $\Delta B_L = - \Delta L_L$.  The different numerical
coefficient arises because there is a 
factor counting the degrees of freedom in the Boltzmann equations which
ultimately leads to the equilibrium condition of
Eq.~(\ref{eq:3bplusL}).  If $(B-L)$ was generated solely for left
handed fermions, and {\em if} it remained solely with left handed fermions,
we would have by Eq.~\eqref{eq:3bplusL}
\begin{equation}
\label{eq:bl1}
B = B_L \approx \frac{1}{4} \, (B-L)\,,
\end{equation}
which incidentally differs from the introductory estimates of
\cite{Harvey:1990qw}
because there, $B_L+L_L=0$ is 
assumed instead of Eq.~(\ref{eq:3bplusL}).  However, there exist
additional interactions in equilibrium --- Yukawa and (possibly)
gaugino interactions --- which flip chirality and convert quarks and
leptons into squarks and sleptons.  Therefore, we need to keep track
of the baryon/lepton content in scalars and right handed fermions,
denoted by $_S$ and $_R$, respectively. The total baryon and lepton
numbers are
\begin{equation}
\label{eq:bl}
B = B_L + B_R + B_S \qquad \qquad L = L_L + L_R + L_S \;.
\end{equation}
Using Eqs.~(\ref{eq:3bplusL}, \ref{eq:bl}), we have
\begin{equation}
B = \frac{1}{4} \left( \, (B-L) + L_R + L_S + 3 B_R +3 B_S \frac{}{} \right)  \;.
\label{eq:intuition}
\end{equation}
Therefore, we see that the deviation of $B$ from $(B-L)/4$ depends on
how much lepton and baryon density is present in scalars and right handed fermions.  These scalar and right handed densities depend sensitively on their masses.  For example, if squark and sleptons are heavy, then their equilibrium densities will be small; the impact of $B_S$ and $L_S$ in Eq.~\eqref{eq:intuition} will be small.  However, if they are light (compared to the temperature at which non-perturbative ${\rm SU}(2)_L$ processes fall out of equilibrium), their equilibrium densities -- and their impact in Eq.~\eqref{eq:intuition} -- may be significant.

\section{\label{sec:equilib}Equilibrium considerations}

In this section, we derive explicit expressions relating $B$ to (a)
$B-L$ in the situation in which the lepton chemical potentials of different flavors
are in equilibrium, or (b) $B/3-L_i$ in the situation in which they are not in equilibrium.  

Consider the physical situation after the freeze out of $B-L$ or
$B/3-L_i$ and before the electroweak phase transition at $T=T_c$.  As
always, Boltzmann equations govern the chemical potentials of all the
particle species.  The reaction rates for the Boltzmann equations are
temperature dependent and the chemical potentials will adjust
themselves depending on the strength of interactions rates.  What is
important for the baryon asymmetry is the time period close to the
electroweak phase transition, the completion of which will effectively
shut off the $B+L$ violating reactions.  Close to this time period at
$T \approx 100$ GeV, the Hubble expansion rate is $H \sim
10^{-14}$ GeV.  Hence, even a very small interaction (e.g. those
suppressed by small Yukawa couplings) can be in equilibrium.

We denote the chemical potential of a particle $X$ by $\mu_X$.
Under the assumption of kinetic equilibrium and for small $\mu_X/T$,
the relation
\begin{equation}
\label{muQ}
X=\frac{T^2}6 g_X k_X\left(\frac{m_X}T\right)\mu_X
\,,
\end{equation}
between the charge density and the chemical potential holds, where
\begin{equation}
k_X\left(\frac{m_X}{T}\right) = \frac{6}{\pi^2} \int_{m_X/T}^\infty dy
\frac{y \sqrt{y^2-(m_X/T)^2} \exp(y)}{\left( \exp(y) \pm 1 \right)^2}\,.
\end{equation}
The $+$ sign holds here for fermions and $-$ for bosons.
In the massless limit, $k_X(0)$ evaluates to $1$ for fermions and $2$ for bosons. 
This is useful for determining the equilibrium value of the
baryon density $B$. In order to compare with
Ref.~\cite{Harvey:1990qw}, but also for an easier evaluation
of the final result, we do not absorb the factor $g_X$
taking account of
the internal colour and isospin degrees of freedom in our definition
of $k_X$.

In order to calculate $B$, one makes use of the fact that
certain reactions are in chemical equilibrium and that
there are conserved charges. We collectively denote
the arising conditions as equilibrium assumptions.
For the MSSM, they can be stated as follows:
\renewcommand{\theenumi}{{\it \alph{enumi}}}
\renewcommand{\labelenumi}{{\it(\theenumi)}}
\begin{enumerate}
\item
\label{ass:isospin}
Isospin violating interactions mediated by $W^\pm$ bosons
are in equilibrium. This implies that
\begin{eqnarray}
\mu_{Q^i}\!\!\!&=&\!\!\!\mu_{u^i_L}=\mu_{d^i_L}\,,
\quad
\mu_{\widetilde Q^i}=\mu_{\widetilde u^i_L}=\mu_{\widetilde d^i_L}\,,
\\\nonumber
\mu_{L^i}\!\!\!&=&\!\!\!\mu_{\nu^i_L}=\mu_{e^i_L}\,,
\quad
\mu_{\widetilde L^i}=\mu_{\widetilde \nu^i_L}=\mu_{\widetilde e^i_L}\,,
\\\nonumber
\mu_H\!\!\!&=&\!\!\!\mu_{H_1^+}=\mu_{H_1^0}=
\mu_{H_2^+}=\mu_{H_2^0}\,.
\end{eqnarray}
For the last
equality, we have also assumed equilibrium of
Yukawa interactions, as listed below. Note that due to the mass-degeneracy
of particles in an isospin doublet, isospin equilibrium here also implies
the absence of the charge density $T^3$ of weak isospin .
\item
\label{ass:Yukawa}
Yukawa interactions of Standard Model Particles are
in equilibrium,
\begin{eqnarray}
\label{Yukawa}
\mu_{Q^i}+\mu_H=\mu_{u^i}\,,\quad
\mu_{Q^i}-\mu_H=\mu_{d^i}\,,\quad
\mu_{L^i}-\mu_H=\mu_{e^i}\,.
\end{eqnarray}
Note that these conditions are automatically consistent with the
${\rm SU}(3)_C$ non-perturbative chiral flip processes being in equilibrium,
\begin{equation}
\sum\limits_{i=1}^3
\left\{
2 \mu_{Q^i} - \mu_{u^i} - \mu_{d^i}
\right\}
=0\,.
\end{equation}
\item
\label{ass:flavour}
Flavor changing interactions in the baryonic sector
are in equilibrium. This allows us to write
\begin{eqnarray}
\mu_Q\!\!\!&=&\!\!\!\mu_{Q^1}=\mu_{Q^2}=\mu_{Q^3}\,,\quad
\mu_u=\mu_{u^1}=\mu_{u^2}=\mu_{u^3}\,,\quad
\mu_d=\mu_{d^1}=\mu_{d^2}=\mu_{d^3}\,.
\end{eqnarray}
\item
\label{ass:neutral}
As a initial condition,
the Universe is neutral with respect to gauge charges.
This means, all charges associated with commuting generators
of local symmetries are imposed to vanish.
We have already 
implicitly incorporated neutrality with respect to isospin $T^3$, associated
with the diagonal generator of ${\rm SU}(2)_L$, as well as
neutrality with respect to color charges. In addition, we
also demand hypercharge-neutrality,
\begin{equation}
Y=0\,.
\end{equation}
\item
\label{ass:WS}
The ${\rm SU}(2)_L$ non-perturbative process is in equilibrium, implying
\begin{equation}
\sum\limits_{i=1}^3
\left(
3\mu_{Q^i}+\mu_{L^i}\right)=0\,.
\end{equation}
\item
\label{ass:B-L}
There are primordial $B/3-L_i$ asymmetries, which possibly originate
from GUT-baryogenesis or leptogenesis, but which are conserved at 
lower temperatures, such as the electroweak scale.
\item
\label{ass:YukawaTriscalarSUSY}
Yukawa and triscalar interactions involving supersymmetric partners
of SM particles are in equilibrium,
\begin{eqnarray}
\label{YukTriSUSY}
\mu_{\widetilde Q^i}+\mu_H=\mu_{\widetilde u^i}\,,\quad
\mu_{\widetilde Q^i}-\mu_H=\mu_{\widetilde d^i}\,,\quad
\mu_{\widetilde L^i}-\mu_H=\mu_{\widetilde e^i}\,,\\
\nonumber
\mu_{\widetilde Q^i}+\mu_{\widetilde H}=\mu_{u^i}\,,\quad
\mu_{\widetilde Q^i}-\mu_{\widetilde H}=\mu_{d^i}\,,\quad
\mu_{\widetilde L^i}-\mu_{\widetilde H}=\mu_{e^i}\,,\\
\nonumber
\mu_{Q^i}+\mu_{\widetilde H}=\mu_{\widetilde u^i}\,,\quad
\mu_{Q^i}-\mu_{\widetilde H}=\mu_{\widetilde d^i}\,,\quad
\mu_{L^i}-\mu_{\widetilde H}=\mu_{\widetilde e^i}\,.
\end{eqnarray}
\item
\label{ass:superequilibrium}
Chemical equilibrium between particles and their superpartners
(``superequilibrium'') holds,
\begin{eqnarray}
\label{supeq}
\mu_{\widetilde Q^i}\!\!\!&=&\!\!\!\mu_{Q^i}\,,\quad
\mu_{\widetilde u^i}=\mu_{u^i}\,\quad
\mu_{\widetilde d^i}=\mu_{d^i}\,,\\\nonumber
\mu_{\widetilde L^i}\!\!\!&=&\!\!\!\mu_{L^i}\,,\quad
\mu_{\widetilde e^i}=\mu_{e^i}\,,\\\nonumber
\mu_{\widetilde H}\!\!\!&=&\!\!\!\mu_H\,.
\end{eqnarray}
Obviously, these relations may be maintained through the
absorption or emission of gauginos. In addition, as observed in
Ref.~\cite{ourbgenesis}, this assumption follows automatically,
provided the conditions~{\it(\ref{ass:Yukawa})}
and~{\it(\ref{ass:YukawaTriscalarSUSY})} hold.
\end{enumerate}

As far as SM particles are concerned, it is well
known that conditions~{\it(\ref{ass:isospin}--\ref{ass:B-L})}
hold (see, {\it e.g.}~\cite{Harvey:1990qw,Dreiner:1992vm}).
For example, to justify assumption~{\it(\ref{ass:Yukawa})},
we consider the electron Yukawa coupling $h_e$, which
is the smallest within the Standard Model sector.
From the electron mass, we can infer
that the electron Yukawa coupling $h_e$ fulfills
$h_e\stackrel{>}{{}_\sim}2.9\times 10^{-6}$, where equality
would apply to the limiting case $\tan\beta=0$.
We denote the thermally averaged net interaction rate for
the process $e_R^-\leftrightarrow H^0 + e_L^-$ as $\Gamma_{h_e}$,
which is to be compared with the Hubble rate
\begin{equation}
H=\sqrt{\frac{8\pi^3}{90}g_*}\frac{T^2}{m_{\rm Pl}}
\approx 2\times 10^{-14}\, {\rm GeV}\,.
\end{equation}
For the purpose of our estimates, we take here and in the following the
electroweak temperature to be $T=100\,{\rm GeV}$ and the
effective number of degrees of freedom of the MSSM
$g_*=228.75$. We remark that at the given temperature, the latter number depends on the sparticle masses, which is of no concern
for the present estimates. If the thermally net averaged
interaction rate for a Yukawa interaction of strength $h_0=1$
is given by $\Gamma_{h_0}$, then we find
$\Gamma_{h_0}\stackrel{>}{{}_\sim}0.03\,{\rm GeV}$
as a condition for the electron Yukawa
rate $h^2_e \Gamma_{h_0}$
to be larger than $3H$. While the precise value of
$\Gamma_{h_0}$ depends on the Higgs boson mass, we expect it at electroweak temperatures to be of order GeV
(for $m_{H^0}=100\,{\rm GeV}$, $\Gamma_{h_0}=0.46\,{\rm GeV}$),
such that electrons and left-handed neutrinos are in equilibrium.

For a generalization of the equilibrium conditions to
the MSSM, we make use of the
observation that in most of parameter
space, condition~{\it(\ref{ass:superequilibrium})} holds, such that we
can the use a common
chemical potential for particles and sparticles,
$\mu_{X^i}=\mu_{\widetilde X^i}$~\cite{ourbgenesis}.
In order to express the charge densities then in terms
of the common chemical potentials {\it via}
relation~(\ref{muQ}), it is convenient to
introduce the expressions
\begin{equation}
\kappa_{X^i}=k_{X^i}+k_{{\widetilde X}^i}\,,\quad
\kappa_X=\sum\limits_{i=1}^3\kappa_{X^i}
\label{eq:notation1}
\end{equation}
for all species except for the Higgs bosons and Higgsinos, for
which we employ
\begin{equation}
\kappa_H=k_{H_1}+k_{H_2}+2 k_{\widetilde H}\,.
\label{eq:notation2}
\end{equation}
We note that $k_X^i=2$ for a massless boson and $k_X^i=1$ for a massless
fermion, while $k_X^i=0$ for both boson and fermion if their mass
is much larger than $T$.

It has already been realized that if $T$ is much larger than
the sparticle masses, the formula for
converting $(B-L)$ to $B$ which is valid for a non-supersymmetric
model also applies to the supersymmetric case, since the simple
proportionality $\widetilde X^i=2 X^i$ then holds for all species and one can use a common chemical potential~\cite{Dreiner:1992vm}.
In turn, the non-supersymmetric conversion formula is
obviously also valid if $T$ is much smaller than the sparticle masses,
simply because the sparticles physically decouple in that limit.
In this paper, we make use of the fact that
$\mu_{X^i}=\mu_{\widetilde X^i}$ is generically fulfilled
in the MSSM at electroweak temperatures even in the intermediate
regime, where sparticle masses are comparable to $T$. The
resulting conversion formula is then
found to depend on the sparticle masses.

A possible concern about the validity of
assumption~{\it(\ref{ass:superequilibrium})} may be that if the gaugino masses are way larger
than $T$, particles and sparticles might not equilibrate. It turns out that this is not
the case.
If assumption~{\it(\ref{ass:YukawaTriscalarSUSY})} is valid,
particle-sparticle equilibrium~(\ref{supeq})
follows algebraically when combining Eqns.~(\ref{Yukawa})
and~(\ref{YukTriSUSY})~\cite{ourbgenesis}.

In order to
argue why assumption~{\it(\ref{ass:YukawaTriscalarSUSY})}
generically holds, we note that
three-point interactions schematically contribute to the Boltzmann
equations as
\begin{equation}
\dot X_i + 3 H X_i = -h^2\Gamma_{h_0}^{(X_1,X_2,X_3)}\left(
\frac{X_1}{k_1}\pm\frac{X_2}{k_2}\pm\frac{X_3}{k_3}
\right)\,,
\end{equation}
where $\Gamma$ denotes a thermally averaged net interaction rate
as defined in~\cite{Cirigliano:2006wh}.
We have extracted here a coupling
constant $h^2$, which can be a Yukawa or a gauge coupling.
An estimate for a certain three-body interaction to be in 
equilibrium then is\footnote{
More precisely, $X_1$ is in equilibrium if $h^2\Gamma_{h_0}^{(X_1,X_2,X_3)}/{k_1} \stackrel{>}{{}_\sim} 3 H$
and $X_{1,2}$ are in equilibrium. For the MSSM, the validity
of assumption~{\it(\ref{ass:YukawaTriscalarSUSY})} can then be derived
starting from the fact that Standard Model
left- and right handed quarks
and leptons are in equilibrium.
}
\begin{equation}
h^2 R^{(X_1,X_2,X_3)}=h^2\Gamma_{h_0}^{(X_1,X_2,X_3)}\left(
\frac{1}{k_1}+\frac{1}{k_2}+\frac{1}{k_3}
\right)\stackrel{>}{{}_\sim}3H\,.
\end{equation}

\begin{figure}[ht]
\vskip.5cm
\begin{center}
\hskip-.2cm
\input{ratefermi}
\end{center}
\vskip-.5cm
\caption{
\label{figure:fermirate}
\small
Yukawa rates $R^{(f_1,\widetilde S,\widetilde f_2)}$ over
$m_{\widetilde f_2}$. We have taken $T=100\,{\rm GeV}$,
$m_{f_1}=0\,{\rm GeV}$ and $m_{\widetilde S}=100\,{\rm GeV}$ (blue solid),
$m_{\widetilde S}=400\,{\rm GeV}$ (red dotted).
}
\end{figure}
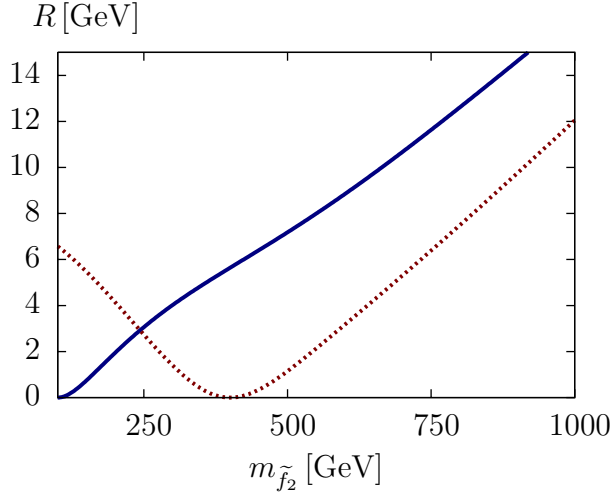

In Figure~\ref{figure:fermirate}, we plot the thermally averaged net interaction
rates for the Yukawa term $f_1 {\widetilde S} {\widetilde f_2}$.
All fields are taken here to be singlets, and factors of particle
of multiplicity can easily be reinserted.
The field $f_1$ is a light
fermion, {\it e.g.} a quark, $\widetilde S$ another heavier fermion,
{\it e.g.} the Higgsino, and $\widetilde f_2$ a scalar field,
{\it e.g.} a squark, the mass of which we vary. When we increase the mass of $\widetilde f_2$, the
equilibration rate $R$ is increasing, because the statistical
factors $k_i$ grow faster than the thermally averaged
net interaction rate $\Gamma$.
Another feature that we see is that
for $m_{\widetilde f_2}=m_{\widetilde S}$ the three-body
process is kinematically forbidden, and the equilibration rate
goes to {\it zero}. We expect that this kinematic blocking is
circumvented through {\it two} by {\it two} processes involving
the additional
emission of a gauge boson from the quark or the squark.

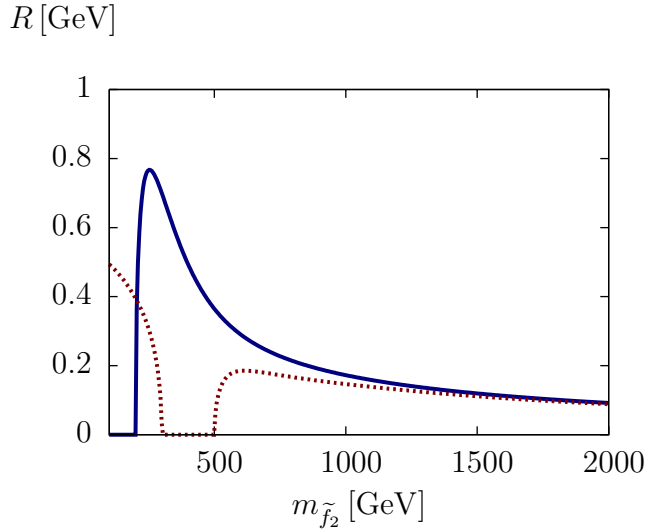
\begin{figure}[ht]
\vskip.5cm
\begin{center}
\hskip-.2cm
\input{ratebose}
\end{center}
\vskip-.5cm
\caption{
\label{figure:boserate}
\small
Triscalar rates $R^{\widetilde f_1,S,\widetilde X_2}$ over
$m_{\widetilde f_2}$. We have taken $T=100\,{\rm GeV}$,
$m_{S}=100\,{\rm GeV}$ and $m_{\widetilde f_1}=100\,{\rm GeV}$ (blue),
$m_{\widetilde f_1}=400\,{\rm GeV}$ (red dotted).
}
\end{figure}

Averaged triscalar rates are plotted in
Figure~\ref{figure:boserate}. We consider the interaction
$\mu \widetilde{f_1}S{\widetilde f_2}$, where all particles
are now scalar and $\mu$ is a triscalar coupling which we take to
be $\mu=100{\rm GeV}$. The fields $\widetilde f_{1,2}$ may
be squarks and $S$ a Higgs boson. We see that even if
we increase the individual masses beyond ${\rm TeV}$ scale,
the equilibration rate remains to be of order ${\rm GeV}$, such
that generically, the triscalar interactions are in equilibrium.
Again, in certain kinematic regions, blocking occurs, which will
be circumvented by gauge boson radiation.

We conclude that at electroweak temperatures,
the equilibration rates $R$ are generically of
order GeV, even for large sparticle masses.
This means that Yukawa and triscalar interactions maintain
chemical equilibrium, even if they are suppressed by couplings
as small as $10^{-6}$.
Since the light quark and lepton fields directly
couple to the Higgs and sparticle fields through Yukawa
interactions, the latter fields are therefore in
chemical equilibrium, even if they are heavy.
By the argument given in
Ref.~\cite{ourbgenesis}, this
implies the chemical equilibrium of particles and sparticles, even
if gauginos are very heavy.

\subsection{Effective lepton flavor equilibrium}

We now discuss the case where it is sufficient to consider
common lepton chemical potentials $\mu_{L}$ and $\mu_{e}$, rather than
the individual $\mu_{L^i}$ and $\mu_{e^i}$. This is relevant in the following two situations.
First, there could be lepton-flavor violating interactions in equilibrium,
which according to the flavor violation in the baryon sector ensure
\begin{equation}
\mu_L=\mu_L^1=\mu_L^2=\mu_L^3\,,\quad
\mu_e=\mu_e^1=\mu_e^2=\mu_e^3\,.
\end{equation}
In this case, the $B/3-L^i$ are no longer conserved, but
$B-L=\sum_{i=1}^3(B/3-L^i)$ still is.
Second, there could be the approximate equalities
$\kappa_{L^i}=\kappa_{L^j}$ and $\kappa_{e^i}=\kappa_{e^j}$
for all $i$ and $j$. Then, we can make use of
$\sum_{i=1}^3\kappa_L^i\mu_{L^i}=\frac 13 \kappa_L\sum_{i=1}^3\mu_{L^i}$
and similarly for $\mu_e$,
such that in the following discussion, it is understood that
$\mu_L=\frac13\sum_{i=1}^3\mu_{L^i}$ and
$\mu_e=\frac13\sum_{i=1}^3\mu_{e^i}$. For example, this
is the situation in the non-supersymmetric models discussed
in Ref.~\cite{Harvey:1990qw}.

Making use of
assumptions~{\it(\ref{ass:isospin}--\ref{ass:flavour},~\ref{ass:YukawaTriscalarSUSY})}
and generalizing the calculation of Ref.~\cite{Harvey:1990qw},
we note that the baryon number $B$, lepton number $L$ and
hypercharge $Y$ are
\begin{eqnarray}
B\!\!\!&=&\!\!\!
\sum\limits_{i=1}^3
\frac 13
\left\{
Q^i+ \widetilde Q^i + u_R^i + \widetilde u_R^i+ d_R^i+ \widetilde d_R^i
\right\}\\\nonumber
\!\!\!&=&\!\!\!
\frac{T^2}6
\sum\limits_{i=1}^3
\left\{
2\kappa_{Q^i}\mu_{Q^i}+ \kappa_{u^i}\mu_{u^i}
+\kappa_{d^i}\mu_{d^i}
\right\}
=
\frac{T^2}6
\left\{
\left[
2 \kappa_{Q}+ \kappa_{u} + \kappa_{d}
\right]\mu_Q
+\left[
\kappa_u-\kappa_d
\right]\mu_H
\right\}
\,,\\
L\!\!\!&=&\!\!\!
\sum\limits_{i=1}^3
\left\{
L^i+ \widetilde L^i + e^i + \widetilde e^i
\right\}\\\nonumber
\!\!\!&=&\!\!\!
\frac{T^2}6
\sum\limits_{i=1}^3
\left\{
2\kappa_{L^i}\mu_{L^i}+ \kappa_{e^i}\mu_{e^i}
\right\}
=
\frac{T^2}6
\left\{
\left[
2\kappa_L+\kappa_e
\right]\mu_L
-\kappa_e \mu_H
\right\}
\,,\\
\end{eqnarray}
\begin{eqnarray}
Y\!\!\!&=&\!\!\!
\sum\limits_{i=1}^3
\left\{
\frac 16 Q^i + \frac 16 \widetilde Q^i
+ \frac 23 u^i + \frac 23 \widetilde u^i
- \frac 13 d^i -\frac 13 \widetilde d^i
- \frac 12 L^i - \frac12 \widetilde L^i
- e^i - \widetilde e^i
\right\}+H
\\\nonumber
\!\!\!&=&\!\!\!
\frac{T^2}6
\sum\limits_{i=1}^3
\left\{
\kappa_{Q^i}\mu_{Q^i}+ 2 \kappa_{u^i}\mu_{u^i}
- \kappa_{d^i}\mu_{d^i} -\kappa_{L^i}\mu_{L^i}- \kappa_{e^i}\mu_{e^i}
\right\}+\frac{T^2}6\kappa_H\mu_H
\\\nonumber
\!\!\!&=&\!\!\!
\frac{T^2}6
\left\{
\left[
\kappa_Q+2\kappa_u-\kappa_d
\right]\mu_Q
+\left[
2\kappa_u+\kappa_d+\kappa_e
\right]\mu_H
-\left[
\kappa_L+\kappa_e
\right]\mu_L
+\kappa_H\mu_H
\right\}\,.
\end{eqnarray}
Note that since we assume $T^3=0$ and $Q=T^3+Y$, where $Q$ is electric 
charge, the condition $Y=0$ is identical to the condition $Q=0$, which is
made in Ref.~\cite{Harvey:1990qw}.

Provided the
assumptions~{\it(\ref{ass:isospin}--\ref{ass:flavour},~\ref{ass:YukawaTriscalarSUSY})}
taken above hold, we see that
we are left with the three chemical potentials
$\mu_Q$, $\mu_L$ and $\mu_H$ after elimination of variables.
Using assumptions~{\it(\ref{ass:neutral})} and~{\it(\ref{ass:WS})}, we can eliminate
$\mu_Q$ and $\mu_H$, while the value of $B-L$ according to
point~{\it(\ref{ass:B-L})} sets the scale of the solution to the homogeneous
system of equations.

Introducing the combinations
\begin{subequations}
\begin{eqnarray}
a_B\!\!\!&=&\!\!\!2\kappa_Q+\kappa_u+\kappa_d\,,\\
a_L\!\!\!&=&\!\!\!-6\kappa_L-3\kappa_e\,,\\
r\!\!\!&=&\!\!\!\kappa_Q+2\kappa_u-\kappa_d+3\kappa_L+3\kappa_e\,,\\
d\!\!\!&=&\!\!\!2\kappa_u+\kappa_d+\kappa_e+\kappa_H\,,
\end{eqnarray}
\end{subequations}
we can express
\begin{subequations}
\begin{eqnarray}
B\!\!\!&=&\!\!\!-\frac{T^2}6 \frac{\mu_L}3
\left\{a_B+(\kappa_d-\kappa_u)\frac rd
\right\}\,,\\
L\!\!\!&=&\!\!\!-\frac{T^2}6 \frac{\mu_L}3
\left\{a_L+\kappa_e\frac rd\right\}\,.
\end{eqnarray}
\end{subequations}
The main result immediately follows,
\begin{equation}
B=\frac{a_B d +(\kappa_d-\kappa_u) r}
{
(a_B-a_L) d + (\kappa_d-\kappa_u-\kappa_e) r
}
(B-L)\,.
\label{eq:leptonflavornonconserved}
\end{equation}
In the limit where all sparticles are superheavy, $m_{\widetilde X}\gg T$,
we recover the result from Ref.~\cite{Harvey:1990qw}. This also holds for
the case when all sparticles are mass degenerate.

\subsection{No lepton flavor equilibrium}
We relax now the assumption of lepton-flavor equilibrium. The
calculation goes along the lines of the lepton-flavor degenerate
case, except that we now have three separate approximately conserved
charges $B/3-L_i$.  The result can be expressed as
\begin{equation}
B=- \frac19 (2\kappa_Q+\kappa_u+\kappa_d)\Lambda
+
\frac{\kappa_u-\kappa_d}
{2\kappa_u+\kappa_d+\kappa_e+\kappa_H}
\Omega\,.
\label{eq:separatecharges}
\end{equation}
where
\begin{subequations}
\begin{eqnarray}
\Lambda\!\!\!&=&\!\!\!
\frac{\alpha\Xi-\varrho\Upsilon}{\alpha\sigma-\beta\varrho}\,,\\
\Omega\!\!\!&=&\!\!\!
\frac{\beta\Xi-\sigma\Upsilon}{\beta\varrho-\alpha\sigma}\,,
\end{eqnarray}
\begin{eqnarray}
\alpha\!\!\!&=&\!\!\!\sum\limits_{i=1}^3
\frac{\kappa_{L^i}+\kappa_{e^i}}{2\kappa_{L^i}+\kappa_{e^i}}
\frac{\kappa_{e^i}+\frac13(\kappa_u-\kappa_d)}
{2\kappa_u+\kappa_d+\kappa_e+\kappa_H}
-1\,,\\
\beta\!\!\!&=&\!\!\!
\frac19(\kappa_Q+2\kappa_u-\kappa_d)
-\frac{1}{27}\sum\limits_{i=1}^3
\frac{\kappa_{L^i}+\kappa_{e^i}}{2\kappa_{L^i}+\kappa_{e^i}}
(2\kappa_Q+\kappa_u+\kappa_d)\,,\\
\varrho\!\!\!&=&\!\!\!
\sum\limits_{i=1}^3
\frac{1}{2\kappa_{L^i}+\kappa_{e^i}}
\frac{\kappa_{e^i}+\frac13(\kappa_u-\kappa_d)}
{2\kappa_u+\kappa_d+\kappa_e+\kappa_H}\,,\\
\sigma\!\!\!&=&\!\!\!-1-
\frac{1}{27}
\sum\limits_{i=1}^3
\frac{1}{2\kappa_{L^i}+\kappa_{e^i}}
(2\kappa_Q+\kappa_u+\kappa_d)\,,\\
\Xi\!\!\!&=&\!\!\!
\sum\limits_{i=1}^3
\frac{1}{2\kappa_{L^i}+\kappa_{e^i}}
\left(\frac B3 -L_i\right)\,,\\
\Upsilon\!\!\!&=&\!\!\!
\sum\limits_{i=1}^3
\frac{\kappa_{L^i}+\kappa_{e^i}}{2\kappa_{L^i}+\kappa_{e^i}}
\left(\frac B3 -L_i\right)\,.
\end{eqnarray}
\end{subequations}

\section{\label{sec:discuss}Discussion of the main results}

We now explore the consequences of
Eqs.~(\ref{eq:leptonflavornonconserved}) and
(\ref{eq:separatecharges}), which are the main results of this paper.
In particular, we give examples of the conversion factors for
mass-spectrum scenarios which are usually considered in models of SUSY
breaking as well as extreme cases, leading to maximal suppression or
enhancement of the conversion factor.  Note that as far as the summed
factors of Eqs.~(\ref{eq:notation1}) and (\ref{eq:notation2}) are
concerned, the kinematic parameters must be in the range $3 \leq
\kappa_X \leq 9$ except for that of the Higgs, for which $2 N_h \leq
\kappa_H \leq 3 N_h $ where $N_h$ is the number of Higgs doublets
($N_h=2$ for the MSSM which is the main illustrative model used in
this paper).  When all the sparticles are much heavier than the
electroweak phase transition temperature and all the SM particles are
light, $\kappa_{X \neq H}=3$ ($\kappa_H=2 N_h$) while when all the
sparticles are light, the other extreme value is taken.  For
$\kappa_{X^i}$ which has an unsummed family index, the values goes
between $1$ (heavy sparticles and light SM particles) and $3$ (light
sparticles and SM particles).

\begin{figure}[ht]
\vskip.5cm
\begin{center}
\hskip-.2cm
\epsfig{file=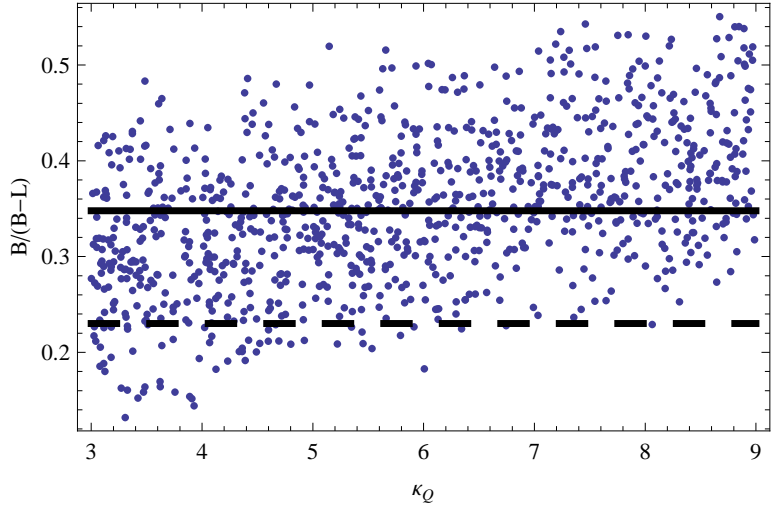, width=4.0in}
\end{center}
\vskip-.5cm
\caption{
\label{fig:flavorequilibscatter}
\small
We plot $B/(B-L)$ as a function of $\kappa_{Q}$ with other
$\vec{\kappa}$ components chosen at random with a flat distribution
between their maximum values.  The solid line corresponds to $8/23$, a
value which is often cavalierly used in the literature and the dashed
line corresponds to the typical mSUGRA value of $38/167$.  Clearly,
the baryon asymmetry is enhanced as more left handed squarks become lighter.
}
\end{figure}

First, consider the lepton flavor equilibrium case of
Eq.~(\ref{eq:leptonflavornonconserved}).  We define the vector $\vec{\kappa}\equiv (\kappa_Q,\kappa_u,\kappa_d,\kappa_L,\kappa_e,\kappa_H)$ and give the following
examples:
\begin{itemize}
\item
SM, which has no sparticles and only one Higgs doublet:
$\vec{\kappa}=(3,3,3,3,3,2)$.
\begin{equation}
B=\frac{28}{79}(B-L)\approx 0.35 (B-L)
\label{eq:ordinarysm}
\end{equation}
\item
Non-supersymmetric two Higgs doublet model: $\vec{\kappa}=(3,3,3,3,3,4)$.
\begin{equation}
B=\frac{8}{23}(B-L)\approx 0.35 (B-L)\,.
\label{eq:standardmssm}
\end{equation}
This is the value usually assumed in the literature for supersymmetric
models, and its usage is sometimes erroneous.
\item
Moderately light sleptons
(e.g. $\tilde{m}_{L^1, \; e^1}/T_c=1.6, \; \tilde{m}_{L^2,\;e^2}/T_c=1.3, \;
\tilde{m}_{L^3,e^3}/T_c=1.1$) in MSSM (generic mSUGRA example):
$\vec{\kappa}=(3,3,3,6,6,4)$.
\begin{equation}
B=\frac{38}{167}(B-L)\approx 0.23 (B-L)\,.
\label{eq:typicalmSUGRA}
\end{equation}
The baryon asymmetry in this generic scenario is therefore
only $2/3$ of what is usually assumed.
\item The largest extreme value: $\vec{\kappa}=(9,3,9,3,3,4)$.
\begin{equation}
B \approx 0.606 (B-L)\,.
\end{equation}
The baryon asymmetry is nearly 3 times that of Eq.~(\ref{eq:typicalmSUGRA}).
It turns out that the enhancement is most sensitive to $\kappa_Q$,
i.e. it becomes large when left handed squarks are light.

\item The smallest extreme value: $\vec{\kappa}=(3,9,3,9,9,4)$.
\begin{equation}
B \approx 0.079 (B-L)\,.
\end{equation}
Note that this $\vec{\kappa}$ corresponds to having light right handed
up squarks, left handed sleptons, and right handed selectrons.  It is also
nearly ``orthogonal'' to the vector of the maximum value case, except
for the $\kappa_H$ entry.
\end{itemize}
As a summary of the lepton flavor equilibrium case, see
Fig.~\ref{fig:flavorequilibscatter} to see the distribution of
$B/(B-L)$ values.  Since the purpose of this plot is to illustrate the
range of kinematic possibilities, we do not analyze the
phenomenological viability of every point in the scatter plot of
Fig.~\ref{fig:flavorequilibscatter} and make appropriate cuts.

\begin{figure}
\vskip.5cm
\begin{center}
\hskip-.2cm
\epsfig{file=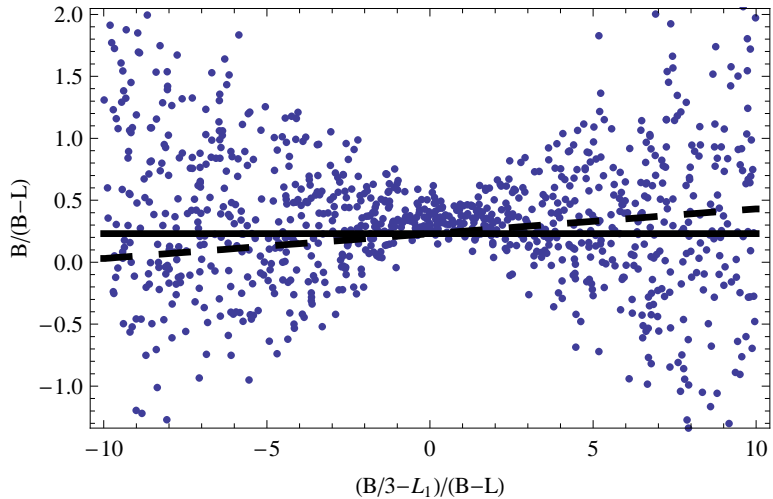, width=4.0in}
\end{center}
\vskip-.5cm
\caption{
\label{fig:flavornonequilib}
\small We plot $B/(B-L)$ as a function of $B/3-L_1$ with other
$\vec{P}$ components chosen at random with a flat distribution between
their maximum values, except for $B/3-L_3$ which has been set to zero.
The solid line corresponds to the value $0.23$, valid for a
typical mSUGRA model, and the dashed line corresponds to
Eq.~(\ref{eq:msugralinearcurve}), the mSUGRA situation
with two flavors contributing to $B-L$.  Note that $B/(B-L)$ can go
negative even when $(B/3-L_1)/(B-L)\sim \mathcal{O}(1)$.}
\end{figure}

Next, consider the lepton flavor violation case of
Eq.~(\ref{eq:separatecharges}).  In this case, $B/(B-L)$ is sensitive
to the magnitude and the signs of $B/3-L_i$ in addition to the
$\kappa_X$.  Hence, we have $B/(B-L)$ being a function of $\vec{P}$ defined as
$\vec{P} \equiv
(\kappa_Q,\kappa_u,\kappa_d,\kappa_{L1},\kappa_{L2},\kappa_{L3},\kappa_{e1},
\kappa_{e2},\kappa_{e3},\kappa_H, B/3-L_1, B/3-L_2, B-L)$.
\begin{itemize}

\item All sparticles heavy: $\vec{P}=
  (3,3,3,1,1,1,1,1,1,2 N_h,B/3-L_1,B/3-L_2,B-L)$.
\begin{equation}
B=\frac{24 + 4 N_h}{66 + 13 N_h} (B-L)\,,
\end{equation} 
which is independent of individual $B/3-L_i$ and depends only on the
sum $B-L$.  It is also identical to Eqs.~(\ref{eq:ordinarysm})  and
(\ref{eq:standardmssm}), and the result of \cite{Harvey:1990qw}.

\item Moderately light sleptons
(e.g. $\tilde{m}_{L^1,\;e^1}/T_c=1.6,\;\tilde{m}_{L^2,\;e^2}/T_c=1.3,\;
\tilde{m}_{L^3,\;e^3}/T_c=1.1$) in MSSM (generic mSUGRA example):
$\vec{P}= (3,3,3,1.85,2.01,2.13,1.85,2.01,2.13,4,B/3-L_1,B/3-L_2,B-L)$.
\begin{equation}
B \approx 0.22 (B-L) + 0.03 (B/3-L_1) + 0.01 (B/3-L_2)\,,
\label{eq:leptonflavormsugra}
\end{equation}
where care has been taken in preserving the precision such that the
small corrections are not merely artifacts of the numerical
truncation.  It is clear that in this case, the result is very similar
to Eq.~(\ref{eq:typicalmSUGRA}) unless $(B/3-L_{1,2})$ is very
different from $B-L$.  In the limit that $L_1=L_2=L/3$,
Eq.~(\ref{eq:typicalmSUGRA}) is recovered.
\item Suppose $B/3-L_3=0$ such that $B/3-L_2= (B-L)-(B/3-L_1)$.  Then
  Eq.~(\ref{eq:leptonflavormsugra}) becomes
\begin{equation}
B \approx 0.23 (B-L) +0.02 (B/3-L_1)\,.
\end{equation}
If we let $r\equiv (B/3-L_1)/(B-L)$, we have
\begin{equation}
B \approx [0.23 +0.02 r] (B-L)
\label{eq:msugralinearcurve}
\end{equation}
Hence, even in the mSUGRA type of spectrum, the baryon number sign can
flip relative to the $B-L$ sign, even if just two flavors contribute to
the lepton asymmetry
with opposite signs and fortuitously cancel with a fine tuning of about
0.1 (i.e. when $r \leq -12$).

\begin{figure}[ht]
\vskip.5cm
\begin{center}
\hskip-.2cm
\epsfig{file=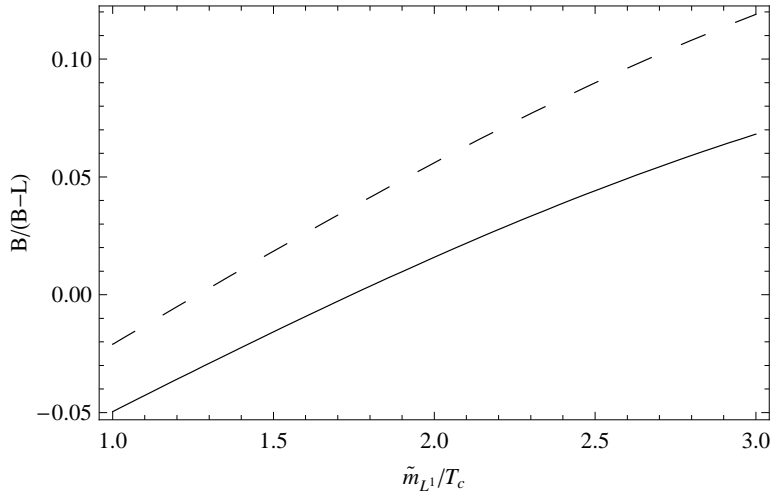, width=4.0in}
\end{center}
\vskip-.5cm
\caption{
\label{fig:signflip}
\small We plot $B/(B-L)$ as a function of $\tilde{m}_{L^1}/T_c$ with only
light sparticles being $\tilde{m}_{u^n}/T_c \approx 4-n$, and
approximately conserved charges fixed to $B/3-L_1 = 2 (B-L)$,
$B/3-L_2=-(B-L)$ and $B/3-L_3=0$.  The dashed curve corresponds to
$\tilde{m}_{e^1}/T_c \approx 2$ and the solid curve corresponds to
$\tilde{m}_{e^1}/T_c \approx 1$.  This demonstrates that sign flip of
$B/(B-L)$ can occur without an unrealistically light sparticle
spectrum or tuning of $B/3 -L_i$.}
\end{figure}

\item A sign flip between $B-L$ and $B$ can be attained for example by
  the following mass spectrum which does not contain unrealistically
  many light sparticles and unrealistic cancellation of $B/3-L_i$:
  $\tilde{m}_{u^n}/T_c \approx 4-n$, and $\tilde{m}_{e^1}/T_c \approx
  1$, $\tilde{m}_{L^1}/T_c\approx 1$ with $B/3-L_1 = 2 (B-L)$,
  $B/3-L_2=-(B-L)$ and $B/3-L_3=0$ gives $\vec{P}=(9,9,3, 3,1,3, 3,1,1, 4, 2
  (B-L), -(B-L),B-L)$ and
\begin{equation}
B\approx -0.05 (B-L).  
\end{equation}
The dependence of this on one of the more sensitive mass parameters
$\tilde{m}_L/T_c$ can be seen in Fig.~\ref{fig:signflip}.

\item A non-vanishing $B$ can be attained even with $B-L=0$.  For
  example, if we choose $B/3-L_1=-(B/3-L_2)$ and $B/3-L_3=0$, we have
  $B-L=0$ but $B\approx -0.08 (B/3-L_1)$ if we take $\tilde{m}_{e^1}/T_c
  \approx 1$.

\end{itemize}
To summarize the range of $B/(B-L)$ that can be attained for
$B/3-L_3=0$, we make a scatter plot of $B/(B-L)$ as a function of
$B/3-L_1$ in Fig,~\ref{fig:flavornonequilib} marginalizing over the remaining free parameters of
$\vec{P}$.  Note that negative values of $B/(B-L)$ can be achieved
without having a huge ratio of $(B/3-L_1)/(B-L)$.

In gravity mediated SUSY breaking scenarios, an upper bound on the
reheating temperature $T_{\textrm{RH}}$ exists due to gravitino decay
effects on big bang nucleosynthesis which can be in conflict with a
successful leptogenesis scenario
\cite{Kawasaki:2004yh,Khlopov:1984pf,Ellis:1984eq}:
i.e. $(B-L)_{\textrm{max}} = c_1 T_{\textrm{RH}} < c_1 T_{\textrm{max}}$
where $c_1 $ is a constant and if we define $B= c_2 (B-L)$ where $c_2$
is the constant that is the focus of this paper, we have $
B_{\textrm{observed}}/|c_2 c_1| < T_{\textrm{RH}}< T_{\textrm{max}}$.
Hence, for those situations in which there is an enhancement of $c_2$,
the squeeze on $T_{\textrm{RH}}$ can be relaxed.  Although
Fig.~\ref{fig:flavornonequilib} looks naively as if a large
enhancement can be achieved through the sparticle kinematic effects,
because such large enhancement cases appear when there is a fine tuned
cancellation between two $B/3 -L_i$, such situations are unlikely to
occur in realistic models.  Milder enhancements (factor of 2 or 3) are
however possible, and hence the lower bound on the reheating
temperature can accordingly be reduced.

We have seen that the conversion factor may take a wide range of values,
depending on the particular sparticle spectrum. Therefore, it is
desirable to present a rule of thumb which applies at least to
parametric scenarios that lead to a successful dark matter genesis
and rely on minimal assumptions of parameters imposed at the
Grand Unified scale. If we focus on the mSUGRA-inspired
scenarios in Ref.~\cite{Allanach:2002nj}, one of the
simplifying assumptions for the Grand Unified scale is a common
value for the squark and slepton masses $m_0$. The renormalization
group running down to the ${\rm TeV}$ scale induces squark masses, that are typically
large enough for $k_{\widetilde Q}^i=k_{\widetilde u}^i=k_{\widetilde d}^i=0$
to be a good approximation. The running of the slepton masses is less strong,
such they may still be close to the electroweak scale. Among the
sleptons, the running is typically
more pronounced for the left handed particles, such that they are
heavier than the right-handed ones. Finally, the bias due to the
Yukawa couplings is less strong for leptons than for baryons, such that
the case of approximate
mass degeneracy for the different flavors is typical.

\begin{table}[ht]
\begin{center}
\begin{tabular}{|l|l|l|l|l|l|l|l|}\hline
\begin{picture}(50,20)(0,0)
      \put(35,10){$m_{\widetilde L}$}
      \put(-6,20){\line(5,-2){62}}
      \put(0,0){$m_{\widetilde e}$}
    \end{picture}
&
        150 GeV&200 GeV&250 GeV&300 GeV&350 GeV&400 GeV\\
\hline
100 GeV & 0.23 & 0.25 & 0.26 & 0.28 & 0.29 & 0.30 \\
\hline
150 GeV & 0.24 & 0.25 & 0.27 & 0.28 & 0.29 & 0.30 \\
\hline
200 GeV & 0.24 & 0.26 & 0.27 & 0.29 & 0.30 & 0.31 \\
\hline
250 GeV & 0.24 & 0.26 & 0.27 & 0.29 & 0.30 & 0.31 \\
\hline
300 GeV & 0.24 & 0.26 & 0.28 & 0.29 & 0.30 & 0.31 \\
\hline
350 GeV & 0.24 & 0.26 & 0.28 & 0.29 & 0.31 & 0.32 \\
\hline
\end{tabular}\medskip
\end{center}
\caption{
\label{table:mSUGRA}
The ratio $B/(B-L)$ for massless quarks and leptons, infinitely heavy
squarks, $\kappa_H=4$
and for given flavor-degenerate slepton masses, corresponding
to mSUGRA motivated scenarios.
}
\end{table}

The size of the mass effects discussed within this paper for typical
mSUGRA scenarios can be inferred from Table~\ref{table:mSUGRA}.
We have taken the squarks to be infinitely massive and for the statistical
factor of the Higgs-particles, we have set $\kappa_H=4$, corresponding
to light Higgsinos and two light complex scalars, while the other
two complex scalars are taken to be very heavy. We see that for light
sleptons, $m_{\widetilde L}=150\,{\rm GeV}$ and
$m_{\widetilde e}=100\,{\rm GeV}$, within this set of typical scenarios
the baryon number is suppressed by roughly a factor of $2/3$, while
for heavier sleptons, we approach the value $B/(B-L)\approx0.35$,
which is usually assumed. We note that none of these mSUGRA-motivated
scenarios allows for a strong first order phase transition, such that in
principle, effects of the onset of electroweak symmetry breaking need to
be taken into account~\cite{Khlebnikov:1996vj,Laine:1999wv}, which we defer to future work.

Our analysis in this paper has assumed that the non-perturbative $B+L$
violating reactions involve only the chemical potentials of the left
handed quarks and leptons.  Ref.~\cite{Ibanez:1992aj} has discussed
the situation in which at high temperatures, the $\mu$-term and the
gaugino mass terms may be neglected to enlarge the global symmetry
group to include a combination of Peccei-Quinn and R-symmetry called
$R_2$, such that the charge coupling anomalously to ${\rm SU}(2)_L$ is
$B+L-R_2$, in which case the chemical potential constraints would
change.  However, since we are dealing with electroweak phase
transition temperatures, $R_2$ is broken and the analysis returns to
the usual non-perturbative $B+L$ violating reactions considered in
this paper.

\section{Summary}

We have considered the kinematic effects on $B-L$ to $B$ conversion
coming from the mass of the MSSM sparticles carrying $B$ and $L$.
The contribution of scalars and right handed fermions carrying $B$ and
$L$ to the equilibrium baryon asymmetry can reduce or
enhance $B$ relative to the standard values used in the literature. 
Explicit formulae for $T\geq T_c$ are given in
Eqs.~(\ref{eq:leptonflavornonconserved}) and
(\ref{eq:separatecharges}).  The typical correction 
for an mSUGRA scenario compared to the usual
values used in the literature is around a factor of $2/3$, but in
some cases, the correction can be dramatic and can even lead to a flip
in the sign between $B$ and $B-L$.  Enhancements of $B$ are also
possible, leading to a mild relaxation of the reheating temperature
bounds coming from gravitino constraints.

\section*{Acknowledgments}
We thank M.~Ramsey-Musolf for discussions and collaboration on a
related project which led to this work.  We thank the hospitality of
KIAS where part of this work was completed.  The work of DJHC and BG is
supported by the DOE Outstanding Junior Investigator Program through
grant DE-FG02-95ER40896. In addition, the work of BG is also
supported by the U.S. Department of Energy
contract No. DE-FG02-08ER41531 and by the Wisconsin Alumni  
Research Foundation.

\end{document}

%% file: ratefermi.tex
\begingroup
  \makeatletter
  \providecommand\color[2][]{%
    \GenericError{(gnuplot) \space\space\space\@spaces}{%
      Package color not loaded in conjunction with
      terminal option `colourtext'%
    }{See the gnuplot documentation for explanation.%
    }{Either use 'blacktext' in gnuplot or load the package
      color.sty in LaTeX.}%
    \renewcommand\color[2][]{}%
  }%
  \providecommand\includegraphics[2][]{%
    \GenericError{(gnuplot) \space\space\space\@spaces}{%
      Package graphicx or graphics not loaded%
    }{See the gnuplot documentation for explanation.%
    }{The gnuplot epslatex terminal needs graphicx.sty or graphics.sty.}%
    \renewcommand\includegraphics[2][]{}%
  }%
  \providecommand\rotatebox[2]{#2}%
  \@ifundefined{ifGPcolor}{%
    \newif\ifGPcolor
    \GPcolortrue
  }{}%
  \@ifundefined{ifGPblacktext}{%
    \newif\ifGPblacktext
    \GPblacktexttrue
  }{}%
  \let\gplgaddtomacro\g@addto@macro
  \gdef\gplbacktext{}%
  \gdef\gplfronttext{}%
  \makeatother
  \ifGPblacktext
    \def\colorrgb#1{}%
    \def\colorgray#1{}%
  \else
    \ifGPcolor
      \def\colorrgb#1{\color[rgb]{#1}}%
      \def\colorgray#1{\color[gray]{#1}}%
      \expandafter\def\csname LTw\endcsname{\color{white}}%
      \expandafter\def\csname LTb\endcsname{\color{black}}%
      \expandafter\def\csname LTa\endcsname{\color{black}}%
      \expandafter\def\csname LT0\endcsname{\color[rgb]{1,0,0}}%
      \expandafter\def\csname LT1\endcsname{\color[rgb]{0,1,0}}%
      \expandafter\def\csname LT2\endcsname{\color[rgb]{0,0,1}}%
      \expandafter\def\csname LT3\endcsname{\color[rgb]{1,0,1}}%
      \expandafter\def\csname LT4\endcsname{\color[rgb]{0,1,1}}%
      \expandafter\def\csname LT5\endcsname{\color[rgb]{1,1,0}}%
      \expandafter\def\csname LT6\endcsname{\color[rgb]{0,0,0}}%
      \expandafter\def\csname LT7\endcsname{\color[rgb]{1,0.3,0}}%
      \expandafter\def\csname LT8\endcsname{\color[rgb]{0.5,0.5,0.5}}%
    \else
      \def\colorrgb#1{\color{black}}%
      \def\colorgray#1{\color[gray]{#1}}%
      \expandafter\def\csname LTw\endcsname{\color{white}}%
      \expandafter\def\csname LTb\endcsname{\color{black}}%
      \expandafter\def\csname LTa\endcsname{\color{black}}%
      \expandafter\def\csname LT0\endcsname{\color{black}}%
      \expandafter\def\csname LT1\endcsname{\color{black}}%
      \expandafter\def\csname LT2\endcsname{\color{black}}%
      \expandafter\def\csname LT3\endcsname{\color{black}}%
      \expandafter\def\csname LT4\endcsname{\color{black}}%
      \expandafter\def\csname LT5\endcsname{\color{black}}%
      \expandafter\def\csname LT6\endcsname{\color{black}}%
      \expandafter\def\csname LT7\endcsname{\color{black}}%
      \expandafter\def\csname LT8\endcsname{\color{black}}%
    \fi
  \fi
  \setlength{\unitlength}{0.0500bp}%
  \begin{picture}(5040.00,3528.00)%
    \gplgaddtomacro\gplbacktext{%
      \csname LTb\endcsname%
      \put(638,660){\makebox(0,0)[r]{\strut{} 0}}%
      \put(638,1007){\makebox(0,0)[r]{\strut{} 2}}%
      \put(638,1354){\makebox(0,0)[r]{\strut{} 4}}%
      \put(638,1702){\makebox(0,0)[r]{\strut{} 6}}%
      \put(638,2049){\makebox(0,0)[r]{\strut{} 8}}%
      \put(638,2396){\makebox(0,0)[r]{\strut{} 10}}%
      \put(638,2743){\makebox(0,0)[r]{\strut{} 12}}%
      \put(638,3090){\makebox(0,0)[r]{\strut{} 14}}%
      \put(1419,440){\makebox(0,0){\strut{} 250}}%
      \put(2502,440){\makebox(0,0){\strut{} 500}}%
      \put(3584,440){\makebox(0,0){\strut{} 750}}%
      \put(4666,440){\makebox(0,0){\strut{} 1000}}%
      \put(2718,110){\makebox(0,0){\strut{}$m_{\widetilde f_2}\, [{\rm GeV}]$}}%
      \put(575,3524){\makebox(0,0)[l]{\strut{}$R\, [{\rm GeV}]$}}%
    }%
    \gplgaddtomacro\gplfronttext{%
    }%
    \gplbacktext
    \put(0,0){\includegraphics{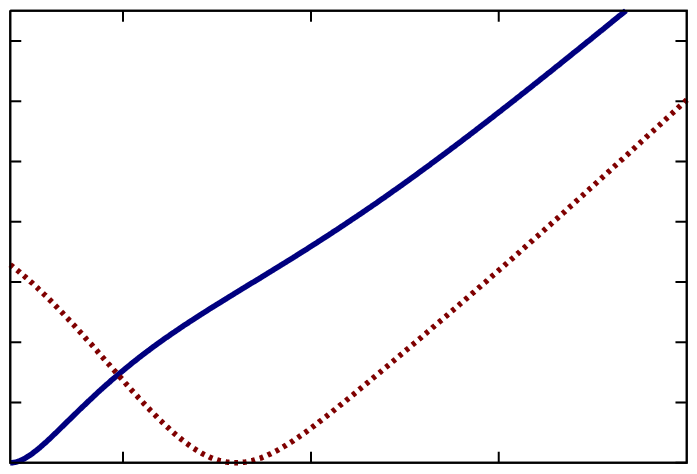}}%
    \gplfronttext
  \end{picture}%
\endgroup

%% file: ratebose.tex
\begingroup
  \makeatletter
  \providecommand\color[2][]{%
    \GenericError{(gnuplot) \space\space\space\@spaces}{%
      Package color not loaded in conjunction with
      terminal option `colourtext'%
    }{See the gnuplot documentation for explanation.%
    }{Either use 'blacktext' in gnuplot or load the package
      color.sty in LaTeX.}%
    \renewcommand\color[2][]{}%
  }%
  \providecommand\includegraphics[2][]{%
    \GenericError{(gnuplot) \space\space\space\@spaces}{%
      Package graphicx or graphics not loaded%
    }{See the gnuplot documentation for explanation.%
    }{The gnuplot epslatex terminal needs graphicx.sty or graphics.sty.}%
    \renewcommand\includegraphics[2][]{}%
  }%
  \providecommand\rotatebox[2]{#2}%
  \@ifundefined{ifGPcolor}{%
    \newif\ifGPcolor
    \GPcolortrue
  }{}%
  \@ifundefined{ifGPblacktext}{%
    \newif\ifGPblacktext
    \GPblacktexttrue
  }{}%
  \let\gplgaddtomacro\g@addto@macro
  \gdef\gplbacktext{}%
  \gdef\gplfronttext{}%
  \makeatother
  \ifGPblacktext
    \def\colorrgb#1{}%
    \def\colorgray#1{}%
  \else
    \ifGPcolor
      \def\colorrgb#1{\color[rgb]{#1}}%
      \def\colorgray#1{\color[gray]{#1}}%
      \expandafter\def\csname LTw\endcsname{\color{white}}%
      \expandafter\def\csname LTb\endcsname{\color{black}}%
      \expandafter\def\csname LTa\endcsname{\color{black}}%
      \expandafter\def\csname LT0\endcsname{\color[rgb]{1,0,0}}%
      \expandafter\def\csname LT1\endcsname{\color[rgb]{0,1,0}}%
      \expandafter\def\csname LT2\endcsname{\color[rgb]{0,0,1}}%
      \expandafter\def\csname LT3\endcsname{\color[rgb]{1,0,1}}%
      \expandafter\def\csname LT4\endcsname{\color[rgb]{0,1,1}}%
      \expandafter\def\csname LT5\endcsname{\color[rgb]{1,1,0}}%
      \expandafter\def\csname LT6\endcsname{\color[rgb]{0,0,0}}%
      \expandafter\def\csname LT7\endcsname{\color[rgb]{1,0.3,0}}%
      \expandafter\def\csname LT8\endcsname{\color[rgb]{0.5,0.5,0.5}}%
    \else
      \def\colorrgb#1{\color{black}}%
      \def\colorgray#1{\color[gray]{#1}}%
      \expandafter\def\csname LTw\endcsname{\color{white}}%
      \expandafter\def\csname LTb\endcsname{\color{black}}%
      \expandafter\def\csname LTa\endcsname{\color{black}}%
      \expandafter\def\csname LT0\endcsname{\color{black}}%
      \expandafter\def\csname LT1\endcsname{\color{black}}%
      \expandafter\def\csname LT2\endcsname{\color{black}}%
      \expandafter\def\csname LT3\endcsname{\color{black}}%
      \expandafter\def\csname LT4\endcsname{\color{black}}%
      \expandafter\def\csname LT5\endcsname{\color{black}}%
      \expandafter\def\csname LT6\endcsname{\color{black}}%
      \expandafter\def\csname LT7\endcsname{\color{black}}%
      \expandafter\def\csname LT8\endcsname{\color{black}}%
    \fi
  \fi
  \setlength{\unitlength}{0.0500bp}%
  \begin{picture}(5040.00,3528.00)%
    \gplgaddtomacro\gplbacktext{%
      \csname LTb\endcsname%
      \put(770,660){\makebox(0,0)[r]{\strut{} 0}}%
      \put(770,1181){\makebox(0,0)[r]{\strut{} 0.2}}%
      \put(770,1702){\makebox(0,0)[r]{\strut{} 0.4}}%
      \put(770,2222){\makebox(0,0)[r]{\strut{} 0.6}}%
      \put(770,2743){\makebox(0,0)[r]{\strut{} 0.8}}%
      \put(770,3264){\makebox(0,0)[r]{\strut{} 1}}%
      \put(1694,440){\makebox(0,0){\strut{} 500}}%
      \put(2685,440){\makebox(0,0){\strut{} 1000}}%
      \put(3675,440){\makebox(0,0){\strut{} 1500}}%
      \put(4666,440){\makebox(0,0){\strut{} 2000}}%
      \put(2784,110){\makebox(0,0){\strut{}$m_{\widetilde f_2}\, [{\rm GeV}]$}}%
      \put(149,3785){\makebox(0,0)[l]{\strut{}$R\, [{\rm GeV}]$}}%
    }%
    \gplgaddtomacro\gplfronttext{%
    }%
    \gplbacktext
    \put(0,0){\includegraphics{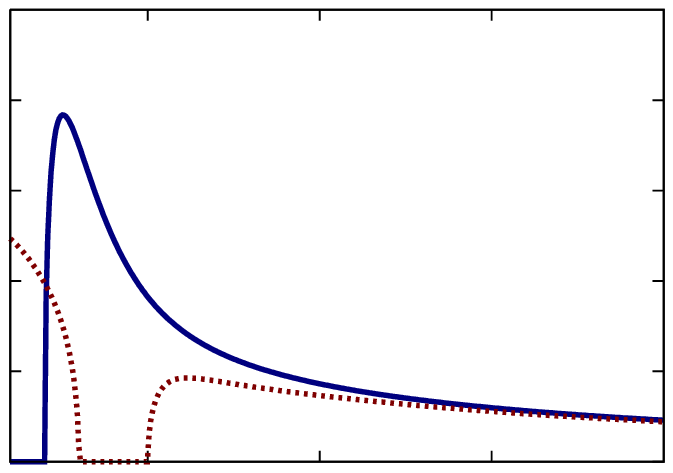}}%
    \gplfronttext
  \end{picture}%
\endgroup